\documentclass[aps,prl,preprint,showpacs,preprintnumbers]{revtex4}
\usepackage{amssymb,latexsym, amsmath,color}

\usepackage{epsfig,epstopdf}

\newcommand{\rem}[1]{}

\newcommand{\bS}{\boldsymbol{S}}
\newcommand{\bm}{\boldsymbol{m}}
\newcommand{\bq}{\boldsymbol{q}}

\newcommand{\bx}{\boldsymbol{x}}

\begin{document}
\title{Geometric order parameter equations} 
\author{Darryl D. Holm$^1$ and Vakhtang Putkaradze$^{2}$ } 
\affiliation{$^1$Department of Mathematics, Imperial College London, SW7
2AZ, UK\\
$^2$Department of Mathematics, Colorado State University, Fort Collins CO 80523 }



\date{\today}
\begin{abstract} \noindent
Aggregation of particles whose interaction potential depends
on their mutual orientation is considered. The aggregation dynamics is derived
using a version of Darcy's law and a variational principle depending on the
geometric nature of the physical quantities. The evolution equations that result 
separate into two classes: either characteristic equations, or gradient
flow equations. We derive analytical solutions of both types of equations which
are collapsed (clumped) states and show their dynamical emergence from smooth
initial conditions in numerical simuations. 

\end{abstract}
\maketitle



\noindent 
{\bf Introduction} 
Many physical processes can be understood as aggregation of individual
`components' at a variety of scales into a final `product'. Diverse examples of
such processes  include the formation of stars, galaxies and solar
systems at large scales, organization of insects and organisms into colonies at
mesoscales and self-assembly of proteins, nanotubes or micro/nanodevices at
micro- and nanoscales  \cite{Whitesides2002}. Some of these processes,
such as  nano-scale self-assembly of molecules are of great
technological interest \cite{Rabani2003,XB2004}.  Of special interest is the case
when the energy of interaction among the assembling particles depends not just
on the distance, but also on mutual orientation. In particular, this Letter is
motivated by recent experiments and simulations on self-assembly of non-circular
particles (squares, hexagons \emph{etc.})
\cite{Boden1999,Grzybowski2001,Grzybowski2004}.  Due to the large number of
particles involved in self-assembly ($10^9-10^{12}$), the development of
continuum descriptions for aggregation or self-assembly is a natural approach
toward theoretical understanding and modeling.  
 \rem{ 
We shall develop a general framework for evolution equation of arbitrary physical
quantity, including orientation-dependent density, and derive evolution
equations. 
  } 
 
Self-assembly at microscopic scales may be simpler than fluid dynamics, because
the Reynolds number is so low that inertia is negligible.  The classic examples of
continuous equations for aggregation are those  of Debye-H\"uckel
\cite{DeHu1923} and   Keller-Segel (KS) \cite{KellerSegel1970}. For a recent
review of developments in this area with an emphasis on biophysical modeling,
see, for example \cite{Ho2003}.  The physics of these models consists of a
conservation law: $\partial_t \rho + \mbox{div} \rho \mathbf{u} = 0$, coupled
with an evolution equation for velocity $\mathbf{u}$ which depends on the density
$\rho$ through a free energy $E$ as  $\mathbf{u}\simeq \mu \nabla \delta
E/\delta \rho$ (velocity proportional to force), in which `mobility' $\mu$ may
also depend on the density. This relation is known as ``Darcy's Law.''  In most
recent works (\cite{HP2005,HP2006,TBL2005} and references within) the energy $E$
is computed under the assumption that the interaction between the particles is
\emph{central}, i.e., that it does not depend on their orientation. This
framework is simple and attractive. However, the self-assembly of some physical
systems may depend on geometric properties such as the mutual orientation of {\em
pieces}. Examples of such systems range from micro-biological (mutual attraction
of cells, viruses or proteins), to electromagnetic (dipoles in continuous media,
orientation of domains), to interactions of living organisms (swarms, herds,
flocks, \emph{etc.}) 

The goal of this Letter is to formulate a procedure to derive and
analyze  evolution equations for systems that self assemble under a flow in
which generalized `velocity' is proportional to generalized `force.'  Although
we aim to derive equations for orientation-dependent motion, we will also show
that the  procedure for obtaining such evolution equations applies generally for
any type of continuum physical quantity. Thus, we obtain a family of
evolution equations for physical quantities such as active scalars, momenta
and fluxes. (A complete list of equations derived by our method is given in
\cite{HP-GOP-2006}.)  After these equations have been formulated, we discuss their
most remarkable feature. Namely, they possess emergent solutions -- or
coherent structures --  in which the self-assembly is {\em singular} (localized
into delta functions). In addition, these localized singular solutions dominate
the long-term dynamics. From the physical point of view, such localized, or
quenched solutions would form the core of the processes of self-assembly and are
therefore of great practical interest. Moreover, the formation of these 
localized solutions is driven by a combination of  nonlinearity and nonlocality
in an evolutionary process that admits an {\em analytical} description of their 
long-term dynamics.

\noindent
{\bf  Geometric evolution equations} 
The general principle for deriving an evolution equation for the self-assembly of
any continuum physical quantity $\kappa$ may be stated as,  ``The local
value of $\kappa$ remains invariant along the characteristic curves of a flow,
whose velocity depends on $\kappa$ through an appropriate Darcy Law.'' This
principle may be formulated in symbols as, $d\kappa(\mathbf{x}(t),t)/dt=0$ along
$d\mathbf{x}/dt=\mathbf{u}[\kappa]$. In the case of particle density $\rho$ in
$n$-dimensional space, for example, the number of particles
$\kappa=\rho\,\mbox{d}^n\mathbf{x}$ has physical meaning. Hence, the time
derivative of $\kappa$ in this case invokes the fundamental chain rule for the
product of the density function times the volume element,
$\kappa=\rho(\mathbf{x}(t),t)\,\mbox{d}^n\mathbf{x}(t)$. Preservation of this
product along
$d\mathbf{x}/dt=\mathbf{u}[\rho]$ yields 
\begin{eqnarray}
(\partial_t \rho
+
\mathbf{u}[\rho]\cdot \nabla \rho
+ \rho\,{\rm div}\,\mathbf{u}[\rho] )\,\mbox{d}^n\mathbf{x}(t)
= 0
\,,
\end{eqnarray}
in which the dependence of the velocity vector $\mathbf{u}[\rho]$ as a
functional of $\rho$ is yet to be determined. As mentioned above, the Darcy Law
approach assumes that velocity $\mathbf{u}$ depends on density $\rho$ through the
gradient of the variation of free energy
$E$ (velocity proportional to `force' with mobility $\mu$). This assumption leads
to the expected continuity equation for density,
\begin{equation} 
\label{cont-eqn}
\partial_t \rho
= -\,{\rm div}\, 
\mu (\rho \nabla \delta E/\delta \rho)^\sharp
\,,
\end{equation} 
where sharp $(\,\cdot\,)^\sharp$ denotes raising the vector index
from covariant to contravariant, so its divergence may be taken.
This gradient-flow method generalizes for other physical quantities with different
geometrical meaning, by noticing that the invariance property 
$d\kappa(\mathbf{x}(t),t)/dt=0$ along $d\mathbf{x}/dt=\mathbf{u}[\kappa]$
may be expressed mathematically as
$\partial\kappa/\partial t + \pounds_{u[\kappa]}\kappa=0$, where
$\pounds_u\kappa$ is the Lie derivative with respect to the vector field
$u=\mathbf{u}\cdot\nabla$ of any geometrical quantity $\kappa$
\cite{MarsdenRatiu}. The key question for the physical modeling is to identify
the analog of Darcy's Law. Namely, what is the vector field $u[\kappa]$ when
$\kappa$ is an arbitrary geometrical quantity?

A surprising clue leading to the geometrical analog of Darcy's Law emerges when
considering the spontaneous appearance of singularities in solutions 
of (\ref{cont-eqn}) for which $\mu$ and $\delta E/\delta \rho$
depend on the {\em average density}, rather than its pointwise value
\cite{HP2005,HP2006}. Those singular solutions obey a weak form of
the continuity equation (\ref{cont-eqn}), expressed by pairing it with an
arbitrary smooth test function $\phi$ and integrating twice by parts, as
\begin{eqnarray}
\label{HP-rho}
\langle \partial_t \rho \,,\, {\phi} \rangle 
&=&
\langle 
\delta\rho \,,\, 
\delta E/\delta \rho 
\rangle 
\quad \mbox{to find} 
\\ 
\delta \rho 
=
-\,{\rm div}\, \rho\mu \nabla \phi
&=&
- \,\pounds_{\mathbf{u}({\phi})}\,\rho
\quad\mbox{so}\quad
\mathbf{u}(\phi)
=
(\mu \nabla \phi)^\sharp
\nonumber
\end{eqnarray}
where spatial integration defines the real-valued pairing
$\langle\,\cdot\,,\,\cdot\,\rangle$ between densities and their dual space of
scalar functions. This expression for the weak solutions of the
continuity equation (\ref{cont-eqn}) provides the clue we seek for expressing
Darcy's Law for the self-assembly of an arbitrary geometric quantity
$\kappa$, not just a density. 

The clue we seek emerges upon re-doing the previous integrations by parts 
slowly and carefully as 
\begin{eqnarray*}
\label{HP-kappa-calc}
\Big\langle \frac{\partial \rho}{\partial t} \,,\, {\phi} \Big\rangle 
&=&
-\,\bigg\langle 
\pounds_{\mathbf{u}({\phi})}\,\rho \,,\, 
\frac{\delta E}{\delta \rho} 
\bigg\rangle 
=:
- \,\bigg\langle 
\mathbf{u}({\phi}) \,,\,\rho \diamond
\frac{\delta E}{\delta \rho} 
\bigg\rangle 
\\ 
&&\hspace{-1cm}  =
-\,\bigg\langle 
{\rm div}\, \rho(\mu \nabla \phi)^\sharp \,,\, 
\frac{\delta E}{\delta \rho} 
\bigg\rangle 
=:-\,
\bigg\langle 
(\mu\diamond\phi)^\sharp
\,,\,
\rho\diamond\frac{\delta E}{\delta \rho} 
\bigg\rangle 
\,.\nonumber
\end{eqnarray*}
Here, we have introduced the \emph{diamond} 
($\diamond$) as the \emph{dual} of the Lie derivative under integration by parts
for any dual pair $(\kappa,b)$ and any vector field $\mathbf{u}$, 
$\langle \kappa \diamond b, \mathbf{u} \rangle=\langle \kappa, 
-\pounds_\mathbf{u} b \rangle$.  This suggests $\mathbf{u}({\phi})=
(\mu \nabla \phi)^\sharp
=-\,(\mu\diamond\phi)^\sharp$. Using
the definition of diamond again gives
\begin{equation}
\big\langle \frac{\partial \rho}{\partial t} \,,\, {\phi} \big\rangle 
=
-\,\big\langle 
(\mu\diamond\phi)^\sharp
\,,\,
\rho\diamond\frac{\delta E}{\delta \rho} 
\big\rangle
=
-\,\big\langle 
\pounds_{(\rho\diamond \frac{\delta E}{\delta \rho})^\sharp }\mu
\,,\,
\phi
\big\rangle 
\label{HP-kappa}
\end{equation}
Replacing $\rho\to\kappa$ here generalizes (\ref{cont-eqn}) to any
quantity $\kappa$ in an arbitrary vector space. We shall derive the particular
versions of this equation for several physical quantities $\kappa$. The result is
a set of novel nonlocal characteristic equations which possess localized 
solutions. Next we shall outline a general procedure for finding those solutions.

\noindent
{\bf Singular solutions for the Geometric order parameter (GOP) equation
(\ref{HP-kappa}).}
Numerical simulations show the development of singular solutions in $\kappa$ for
some types of $\kappa$ (e.g., scalars) and the absence of those singularities for
other $\kappa$'s (e.g., vector fields). These solutions are reminiscent of the
\emph{clumpon}  singularities \cite{HP2005,HP2006}, which dominate the long term
dynamics once they have formed. Thus, we seek a particular  solution of
(\ref{HP-kappa}) expressed as a $\delta$-function parametrized by
coordinate(s) $s$ on a submanifold of ambient space, 
\begin{equation} 
\label{clumpons} 
\kappa(\bx,t)=\int p(s,t) \delta \left(\bx - \mathbf{q}(s,t) \right)\,ds
\,.
\end{equation}
To derive the equations for $p(s,t)$ and $\mathbf{q}(s,t)$, we
substitute (\ref{clumpons}) into  (\ref{HP-kappa}) and integrate the right-hand
side by parts to extract the term proportional to $\kappa$ as follows: 

\begin{equation}
\label{HP-kappa2}
\frac{\partial p}{\partial t} \phi \left(\mathbf{q}(s,t) \right)
+
\frac{\partial \mathbf{q}}{\partial t}\cdot \nabla \phi 
\left(\mathbf{q}(s,t) \right) 
= \,
\bigg\langle 
\kappa,  {\cal N}_\kappa \phi 
\bigg\rangle 
\end{equation}
where ${\cal N}_\kappa$ is a linear operator acting on $\phi$ depending on the nature of $\kappa$. For example, for densities $\kappa = \rho \mbox{d}^3 \bx$, ${\cal N}_\kappa \phi =-\mu \nabla \delta E/\delta \rho \cdot \nabla \phi\left(\mathbf{q}(s,t)\right)$. 
If the right-hand side of (\ref{HP-kappa2}) only contains the function $\phi$ and
its gradient, then singular solutions (\ref{clumpons}) are possible.
However, one must be careful because this condition could be over-determined
and thus the existence of singular solutions of (\ref{HP-kappa}) for the evolution
of an arbitrary geometric quantity might not be guaranteed. Two classes of
geometric quantities admitting singular solutions of (\ref{HP-kappa}) are known
\cite{HP-GOP-2006}. One class includes, for example, scalars, 1-forms and
2-forms, and gives characteristic equations for which the characteristic velocity
is a nonlocal vector function. A second class, which encompasses in particular an 
orientation-dependent density, is a nonlinear nonlocal diffusion equation. In
each case, one must compute the particular expression for the right hand side of
(\ref{HP-kappa}), and integrate by parts to extract the function $\phi$ and its
derivatives.  Since our main interest lies with the second case, we shall
describe  the nonlocal characteristic equations only very briefly. It is
essential to discuss them, however, because of their interesting mathematical and
physical properties and their many possible applications. 

\noindent
{\bf  Nonlocal characteristic equations } 
The fundamental example is for a scalar $\kappa=f$. The evolution of a scalar
by (\ref{HP-kappa}) obeys
\begin{equation}
\frac{\partial f}{\partial t} =
 -\,\pounds_{(f\diamond\frac{\delta E}{\delta f})^\sharp}\mu[f]
= -\,\Big(\frac{\delta E}{\delta f}\nabla \mu[f] \Big)^\sharp\cdot\nabla f
\,.\label{scalareq} 
\end{equation}
Equation (\ref{scalareq}) can be rewritten in characteristic form as 
$df/dt=0$ on 
$
d\mathbf{x}/dt
=
\Big(\frac{\delta E}{\delta f}\nabla \mu[f] \Big)^\sharp.
$ 
The characteristic speeds of this equation are {\em nonlocal}
when $\delta E/\delta f$ and $\mu$  are chosen to
depend on the {\em average value}, $\bar{f}$.

One may verify that the scalar equation (\ref{scalareq}) admits weak solutions
(\ref{clumpons}). Figure~1 
shows the spatio-temporal
numerical evolution of $H*f$ given by (\ref{scalareq}) with initial conditions of
the type (\ref{clumpons}) with  $\delta$-functions whose strengths are random
numbers between $\pm 1/8$. We have taken $\delta E/\delta f=H*f$ where
$H$ is the inverse Helmholtz operator $H(x)=e^{-|x|/\alpha}$ with $\alpha=1$. We
see the evolution of sharp ridges in $\overline{f}=H*f$, which corresponds to
$\delta$-functions in the solutions $f(x,t)$.

Explicit equations for the evolution of strengths $p_a$ and
coordinates $\mathbf{q}_a$ for a sum of $\delta$-functions in (\ref{clumpons}) may
be derived using (\ref{HP-kappa2}) as
\begin{eqnarray}
\frac{\partial p_a(t,s)}{\partial t} 
&=& 
{p}_a(t,s) 
\,{\rm div}\,
\Big(\frac{\delta E}{\delta f}
\nabla \mu[f]\Big)^\sharp
\bigg|_{\bx = \bq_a(t,{s})}
\label{weak-fsoln-peqn}\\
{p}_a(t,s) \, \frac{\partial \mathbf{q}_a(t,s)}{\partial t} 
&=&
{p}_a(t,s)\, 
\Big(\frac{\delta E}{\delta f}
\nabla \mu[f]\Big)^\sharp
\bigg|_{\bx = \bq_a(t,{s})}
\label{weak-fsoln-qeqn}
\end{eqnarray}
for $a=1,2,\dots,N$. A solution containing a single $\delta$-function satisfies
$\dot{p}=-A p^3$, so an initial condition $p(0)=p_0$, evolves
according to $1/p(t)^2=1/p_0^2+2 A t$. For the choice of our parameters in
simulations, $A=2$. The comparison of $1/p^2$ from numerics with this theoretical
prediction is shown in Figure 2. 

We only mention that the evolution of a general 1-form (velocity) and
2-form (flux)  may also be cast into the form of a nonlocal characteristic
equation. For a 1-form in two- or three-dimensional space 
$\mathbf{A} \cdot \mbox{d} \mathbf{x}$, the equation is quite complicated
\cite{HP-GOP-2006}. However, that equation allows a simplification in the case
$\mathbf{A}=\nabla \psi$. The equation for potential $\psi$ reads,
with nonlocal ${\delta E/\delta\psi}$ and $\mu[\psi]$,
\begin{equation}
\frac{\partial \psi }{\partial t}= \Big( \frac{\delta E}{\delta\psi}
\nabla \mu[\psi] \Big)^\sharp
\boldsymbol{\cdot}\nabla \psi
\,.
\label{APsieq}
\end{equation} 
This equation for the potential $\psi$ has the same nonlocal
characteristic structure as the scalar equation (\ref{scalareq}). 

Similarly, the evolution of 2-form fluxes 
$\mathbf{B} \cdot d\bS= B_1 \mbox{d}x^2 \wedge \mbox{d}x^3-B_2 \mbox{d}x^1
\wedge \mbox{d}x^3+ B_3 \mbox{d}x^1 \wedge \mbox{d}x^2$ also simplifies. Again,
the evolution for a general flux $\mathbf{B} \cdot d\bS$ is
quite complicated, but it can be simplified  for the case  ${\rm
div}\,\mathbf{B}=0$ when $\mathbf{B}$ only depends on two coordinates $(x,y)$.
In this case, equation (\ref{HP-kappa}) may be written for the stream function
$\Psi$ in  $\mathbf{B}=\mbox{curl}\,\Psi\mathbf{\hat{z}} 
=\nabla \Psi \times \mathbf{\hat{z}}$ as 
\begin{equation} 
\label{psieqB2} 
\frac{\partial \Psi}{\partial t}
=  \Big(\frac{\delta E}{\delta \Psi} \nabla \Phi \Big)^\sharp \cdot \nabla \Psi
\,, 
\end{equation}
when we choose mobility to be $\mu=\mbox{curl}\,(\mathbf{\hat{z}} \Phi) 
=\nabla \Phi\times\mathbf{\hat{z}} $. 
Choosing ${\delta E/\delta \Psi}$ and $\Phi$ to depend on the average value
$\bar{\Psi}$ again yields a nonlocal characteristic equation. 

\noindent
{\bf Nonlocal aggregation equations} 
Next, we consider the case which is the main focus of the paper, namely the 
orientation-dependent density. The evolution equations arising in this case are
similar to (\ref{cont-eqn}) for mass density. 

Density at each point is given by one number, whereas the number of coordinates
necessary to describe the orientation by an element of the rotation group $SO(n)$
depends on the dimension of the space $n$. In two dimension, one number --
turning angle -- is enough, whereas in three dimensions three numbers are
required. Keeping in spirit with the framework (\ref{HP-kappa}), we see that we
cannot make a theory describing evolution of orientation per se, because $SO(n)$
is not a linear space. Instead, we need to consider the tangent space to $SO(n)$
taken at the identity, commonly denoted $so(n)$, which is a linear space. The
description in terms of coordinates on $so(n)$ is simple and corresponds to
physical intuition.   So, let us define $\kappa$ to be a {\em pair} of densities
$(\rho,\sigma)$  where $\rho$ is mass or number density and
$\sigma$ is orientation density of particles, taking values in $so(n)$.  Hence,
we set $\kappa=\sum \kappa^a e_a$ where $e_a$ are basis vectors spanning 
the space of $\kappa$, $a=0,1,2,3$ with $a=0$ corresponding to the mass density. 
In 3-dimensional space, an elegant description of $\kappa$ exists using
quaternions, obtained by associating the density part $a=0$ with the scalar part
and $a=1,2,3$ with the `vector' part of the quaternion. In two dimensions,
an analogous description uses complex numbers. For orientation-dependent
aggregation equation in 3D equation (\ref{HP-kappa}) reads 
\begin{equation} 
\label{so3eq} 
\frac{\partial \kappa\,^b }{\partial t}
=
-\,\mbox{div}
\Big(   
\mu^b[\overline{\kappa}] 
\Big(\kappa^a\nabla \frac{\delta E}{\delta \kappa^a}
\Big)
\Big) 
,\ 
a,b=0,1,2,3,
\end{equation}
or, explicitly for  $\rho$ and $\sigma^b,\ b=1,2,3$: 
\begin{eqnarray} 
\label{so3eqrho} 
\frac{\partial \rho }{\partial t}
=
-\,\mbox{div}
\Big( \mu_\rho  
\Big(\rho \nabla \frac{\delta E}{\delta \rho} 
+
\sigma^a \nabla \frac{\delta E}{\delta \sigma^a} \Big)
\Big)  \\ 
\label{so3eqsigma} 
\frac{\partial \sigma^b }{\partial t}
=
-\,\mbox{div}
\Big( \mu_\sigma^b
\Big(\rho \nabla \frac{\delta E}{\delta \rho} 
+
\sigma^a \nabla \frac{\delta E}{\delta \sigma^a} \Big)
\Big) 
\end{eqnarray} 
(Summation on $a$ is assumed, but no summation on $b$.) 

Interestingly enough, equations (\ref{so3eqrho},\ref{so3eqsigma}) allow weak
solutions  of the form (\ref{clumpons}). A special condition for
existence of such solutions is that all mobilities $\mu^b$ must be constant along
the flow. Then, even though ansatz  (\ref{clumpons}) leads to an over-determined
system, this system is consistent and  for any real $K$, the following is an
\emph{exact} solution of (\ref{HP-kappa},\ref{clumpons}): 
\begin{equation} 
\label{orientonspq}
p^b=K \mu^b[\overline{\kappa}] 
\hskip0.5cm  
K \frac{\partial q^j}{\partial t}
=  
\Big(p^a \frac{\partial}{\partial x^j} 
\frac{\delta E}{\delta \kappa^a}\Big| _{ \mathbf{r}
=
\mathbf{q}(s,t) } 
\Big)^\sharp. 
\end{equation} 
In Figure 3, 
we present a simulation of the density and
orientation in two dimensions. Starting with density and orientation spread
unequally over two Gaussian `clumps', we end up with a stationary solution whose
density and orientation are concentrated into one clump. This is the nature
of self assembly in this case. 

\noindent
{\bf Further developments} Equation (\ref{HP-kappa}) also encompasses
ideal fluid vorticity dynamics (which is not associated with
Darcy's law).  For the pairing  $\langle \omega \,,\, m \rangle =
\int \boldsymbol{\omega \cdot m} \, dV
$ between a divergenceless vector field
$\omega = \boldsymbol{\omega} \cdot\nabla$ and a one-form
density $m = \bm\cdot d\bx\otimes dV$, set
$
\Big\langle \frac{\partial {\omega}}{\partial t} \,,\, m \Big\rangle=
\Big\langle
\delta{\omega}
\,,\,
\frac{\delta E}{\delta {\omega}}
\Big\rangle
$ 
and choose variation 
$
\delta \omega 
= \pounds_{{\rm curl} m}\,\omega
= \big[{\rm curl} m\,,\,\omega\big]
$. After two integrations by parts, one finds
$\partial_t \omega=
-\,\big[\,{\rm curl}\frac{\delta E}{\delta {\omega}}\,,\,\omega\ \big] $ which
recovers fluid vortex dynamics when ${\rm curl}\frac{\delta
E}{\delta \boldsymbol{\omega}}=\mathbf{u}$ and $\boldsymbol{\omega}={\rm
curl}\mathbf{u}$. It will be interesting to see what other physical systems
remain to be discovered in equation (\ref{HP-kappa}).

\noindent 
{\bf Acknowledgements} 
The authors were partially supported by NSF grant NSF-DMS-05377891. The work of
DDH was also partially supported  by the US Department of Energy, Office of 
Science, Applied Mathematical Research. 

\newpage
\begin{center}
{\bf \large Figure Captions. }
\end{center} 
\begin{itemize} 
\item[Figure 1.]
Numerical simulation of scalar $f$, starting with initial condition which is a
set of $\delta$-functions in $f$. The vertical coordinate represents 
$\overline{f}=H*f$, which remains finite even when $f$ forms $\delta$-functions, where
$H(x)=e^{-|x|}$. 
The horizontal coordinate is space; the vertical coordinate is time.
\item[Figure 2.]
Evolution of the $\delta$-function strength $1/p(t)^2$ versus time (circles).
The theoretical prediction $1/p_0^2+4 t$ is shown as a solid line obtained
without any fitting parameters. 
\item[Figure 3.]
 Numerical simulation of density $\overline{\rho}$
starting with initial conditions for $\rho$ and $\sigma$ that are two
Gaussian clumps of unequal strength 
$(\rho,\sigma)(\mathbf{r},0)=\sum_{clumps} 
(\rho_0,\sigma_0) e^{-\mathbf{r}^2/l_{(\rho \sigma)}^2}$. 
The initial conditions  for $\rho$ and $\sigma$ are similar in shape but different
in width. For this simulation, we have taken $l_\sigma=2 l_\rho$, $H(\mathbf{x})=e^{-|\mathbf{x}|}$.   Top: initial
conditions, bottom: finite solution for $t=30$. Left column: density, right
column: orientation. White represents the domains of high density/orientation.  
\end{itemize} 

\end{document}